\documentclass[fleqn,usenatbib]{mnras}

\usepackage{newtxtext,newtxmath}

\usepackage[T1]{fontenc}
\usepackage{ae,aecompl}

\usepackage{graphicx}	
\usepackage{amsmath}	
\usepackage{amssymb}	
\usepackage{etoolbox}
\usepackage{array}
\makeatletter
\patchcmd\@combinedblfloats{\box\@outputbox}{\unvbox\@outputbox}{}{\errmessage{\noexpand patch failed}}
\makeatother

\title[Ages of pulsar]{Ages of radio pulsar: long-term magnetic field evolution}

\author[Andrei P. Igoshev]{
Andrei P. Igoshev,$^{1}$\thanks{E-mail: ignotur@gmail.com}
\\
$^{1}$Department of Physics, Technion, Haifa 3200003, Israel
}

\date{Accepted XXX. Received YYY; in original form ZZZ}

\pubyear{2015}

\begin{document}
\label{firstpage}
\pagerange{\pageref{firstpage}--\pageref{lastpage}}
\maketitle

\begin{abstract}
We use the Bayesian approach to write the posterior probability density for the three-dimensional velocity of a pulsar and for its kinematic age. As a prior, we use the bimodal velocity distribution found in a recent article by Verbunt, Igoshev \& Cator (2017). When we compare the kinematic ages with spin-down ages we find that in general they agree with each other.
In particular, maximum likelihood analysis sets the lower limit for the exponential magnetic field decay timescale at $8$~Myr with slight preference of $t_\mathrm{dec} \approx 12$~Myr and compatible with no decay at all.  
One of the objects in the study, pulsar B0950+08 has kinematic and cooling ages $\approx 2$~Myr which is in strong contradiction with its spin-down age $\tau\approx 17$ Myr. The 68 per cent credible range for the kinematic age is 1.2--8.0~Myr.  We conclude that the most probable explanation for this contradiction is a combination of magnetic field decay and long initial period. Further timing, UV and X-ray observations of B0950+08 are required to better constrain its origin and evolution.
\end{abstract}




\section{Introduction}

The knowledge of precise radio pulsar ages is important because this parameter helps us to constrain fundamental physical processes in neutron star (NS) such as NS cooling (probing the properties of matter in the core) and the evolution of magnetic fields and obliquity angle (probing the properties of matter in the outer crust and magnetosphere, see examples in \citealt{yakovlev2004, Noutsos2013, 1992ARA&A..30..143C, 2015AN....336..831I, biryukov2017}).
The kinematic age is the ratio of the NS displacement from the Galactic plane (birth location of OB stars which are NS progenitors) to NS vertical speed. The kinematic age is independent of the model for NS interior, which makes it especially useful test for internal pulsar timescales.
Another age estimate, so called spin-down age $\tau = P / (2\dot P)$ where $P$ is the rotational period and $\dot P$ is the period derivative of the pulsar, is strongly sensitive to the magnetic field and obliquity angle evolution as well as to initial NS properties, see e.g. \cite{2014MNRAS.444.1066I}.

For isolated radio pulsars with spin-down ages below $\approx 20$~Myr (quarter of the Galactic vertical oscillation period which is $\approx 87$~Myr according to \citealt{2008gady.book.....B}) the kinematic ages can be estimated unambiguously. Older pulsars could have completed a few oscillation cycles in the Galactic gravitational potential and the probability of different ages splits equally between multiples of the vertical oscillation period. Pulsars with spin-down ages $\tau < 20$~Myr are especially useful to test models of the magnetic field evolution which might occur on a $10$~Myr timescale according to older works by \cite{mf_decay1} and \cite{mf_decay2}. This magnetic field decay timescale was studied in multiple population synthesis \citep{1992A&A...254..198B,vanLeeuwen,vanLeeuwenThesis,1997MNRAS.289..592L,faucher2006}: no convincing evidences were found for it, except a recent work by  \cite{2018arXiv180302397C} who identified timescale of 4~Myr.

The main reasons to re-analyze the kinematic ages (after the work by \citealt{Noutsos2013}) are the new precise measurements for the parallax and proper motion by \cite{Deller2018} and the introduction of a new bimodal velocity distribution derived in \cite*{verbunt2017}. This velocity distribution is significantly different from the earlier used distribution by \cite{hobbs2005} and more similar to earlier estimates derived in \cite*{arzoumanian}. The kinematic ages of radio pulsars are quite sensitive to unmeasurable radial velocity, therefore a use of more precise velocity distribution is important to derive a correct kinematic age estimate. The secondary reason is to introduce proper treatment of uncertainties in distance and proper motion measurements.

Unlike simple analysis of kinematic - spin-down ages diagram criticized in \cite{1997MNRAS.289..592L}, our analysis includes radial velocities, unknown birth location and initial spin-down ages. Unlike the population synthesis approach, our analysis is independent of luminosity function and exact beaming model. 

Among other things, we highlight here the pulsar B0950+08 (alternative name J0953+0755) with kinematic and thermal age $t\approx 2$~Myr which is much smaller than its spin-down age. 
This is expected to be observed if some of NS experiences faster magnetic field evolution or are born with longer initial periods.

The article is structured as follows: in the third Section we derive the posterior distribution for the total velocity and show to what  extent it is sensitive to the unknown radial component; in the fourth Section we introduce an analytical estimate for the kinematic age and elaborate on it to take the following effects into account: uncertain latitudinal velocity, uncertain birth position and distance. At the end of this section we show where this analytical approach is accurate and introduce a proper treatment of the Galactic gravitational potential and perform Markov Chain Monte Carlo simulations for the kinematic age of B0950+08. In the fifth section we describe a maximum likelihood method to estimate the magnetic field decay timescale and show the results.
In the last section we discuss obliquity angle evolution and compare it with new estimates of the kinematic ages. 

\section{Sample}
The primary task of this article is to develop a formalism which can be used further in application to individual objects. For illustrative purposes and to study the effects of the possible magnetic field and obliquity angle evolution we use the same sample as in \cite{verbunt2017} adding measurements from \cite{Deller2018} restricting ourself to objects with $\tau < 20$~Myr moving away from the Galactic plane. It makes 43 objects in total. 

This sample contains the most precise measurements of the parallax and proper motions for isolated radio pulsars available today from works by \cite{2002ApJ...571..906B,2003ApJ...593L..89B,2001ApJ...550..287C,2004ApJ...604..339C,2009ApJ...698..250C,2009ApJ...701.1243D,Deller2018,2015A&A...577A.111K}. All these measurements are performed by means of radio interferometry with very long baseline. The list of pulsar names can be found in Table~\ref{t:res}.

\begin{table}
\caption{The 68 per cent credible intervals for three-dimensional velocity and kinematic age of pulsars
with $\tau < 20$ Myr. The intrinsic accuracy for calculations of the three-dimensional velocity is 12~km~s$^{-1}$}
\label{t:res}
\setlength\extrarowheight{4pt}
\begin{center}
\begin{tabular}{lccc}
\hline
Name        & $v$                       & $t_\mathrm{kin}$     & $\tau$  \\
            & km/s	                    & Myr                  & Myr  \\
\hline
J0055+5117  &  $353_{-35}^{+224}$  &  $2.5_{-0.4}^{+1.2}$  &  3.5    \\
J0102+6537  &  $141_{-12}^{+482}$  &  $3.6_{-1.3}^{+2.2}$  &  4.5    \\
J0108+6608  &  $494_{-12}^{+176}$  &  $0.3_{-0.1}^{+0.1}$  &  1.6    \\
J0139+5814  &  $553_{-94}^{+165}$  &  $0.7_{-0.2}^{+0.2}$  &  0.4    \\
J0358+5413  &  $ 94_{-12}^{+376}$  &  $0.2_{-0.1}^{+0.8}$  &  0.6    \\
J0406+6138  &  $588_{-59}^{+200}$  &  $1.0_{-0.1}^{+0.1}$  &  1.7    \\
J0454+5543  &  $341_{-12}^{+224}$  &  $0.8_{-0.2}^{+0.4}$  &  2.3    \\
J0601-0527  &  $212_{-12}^{+282}$  &  $3.5_{-0.9}^{+1.8}$  &  4.8    \\
J0629+2415  &  $106_{-12}^{+129}$  &  $11.0_{-3.3}^{+4.7}$  &  3.8    \\
J0630-2834  &  $ 94_{-12}^{+400}$  &  $1.5_{-0.6}^{+2.2}$  &  2.8    \\
J0659+1414  &  $ 82_{-12}^{+435}$  &  $0.6_{-0.2}^{+1.2}$  &  0.1    \\
J0729-1836  &  $176_{-12}^{+482}$  &  $0.3_{-0.1}^{+1.3}$  &  0.4    \\
J0826+2637  &  $306_{-12}^{+259}$  &  $2.3_{-0.8}^{+3.5}$  &  4.9    \\
J0922+0638  &  $576_{-59}^{+176}$  &  $1.6_{-0.1}^{+3.0}$  &  0.5    \\
J0953+0755  &  $ 47_{-12}^{+388}$  &  $1.9_{-0.6}^{+5.5}$  &  17.4    \\
J1136+1551  &  $671_{-12}^{+153}$  &  $0.7_{-0.1}^{+3.0}$  &  5.0    \\
J1509+5531  &  $988_{-47}^{+118}$  &  $2.2_{-0.2}^{+2.4}$  &  2.3    \\
J1543-0620  &  $435_{-59}^{+200}$  &  $5.2_{-0.2}^{+8.7}$  &  12.8    \\
J1559-4438  &  $341_{-94}^{+282}$  &  $2.5_{-0.5}^{+0.8}$  &  4.0    \\
J1623-0908  &  $200_{ 24}^{+388}$  &  $2.8_{-0.4}^{+4.4}$  &  7.8    \\
J1645-0317  &  $435_{-12}^{+235}$  &  $5.2_{-0.6}^{+6.3}$  &  3.5    \\
J1703-1846  &  $282_{-12}^{+329}$  &  $4.4_{-1.1}^{+2.2}$  &  7.4    \\
J1735-0724  &  $635_{-59}^{+235}$  &  $4.2_{-0.5}^{+1.8}$  &  5.5    \\
J1741-0840  &  $153_{-12}^{+353}$  &  $9.9_{-2.3}^{+3.5}$  &  14.2    \\
J1820-0427  &  $353_{-35}^{+282}$  &  $1.2_{-0.2}^{+0.3}$  &  1.5    \\
J1833-0338  &  $424_{-59}^{+353}$  &  $0.3_{-0.1}^{+0.2}$  &  0.3    \\
J1840+5640  &  $341_{-12}^{+235}$  &  $3.3_{-1.0}^{+2.9}$  &  17.5    \\
J1901-0906  &  $176_{-12}^{+141}$  &  $13.5_{-6.9}^{+3.7}$  &  17.2    \\
J1913+1400  &  $165_{-12}^{+341}$  &  $6.7_{-2.1}^{+3.0}$  &  10.3    \\
J1919+0021  &  $576_{-71}^{+176}$  &  $2.0_{-0.2}^{+0.3}$  &  2.6    \\
J1932+1059  &  $176_{ 24}^{+506}$  &  $0.2_{-0.1}^{+0.7}$  &  3.1    \\
J1937+2544  &  $224_{-12}^{+294}$  &  $2.8_{-1.0}^{+1.3}$  &  5.0    \\
J2022+2854  &  $176_{-12}^{+247}$  &  $1.7_{-0.5}^{+0.8}$  &  2.9    \\
J2022+5154  &  $106_{-12}^{+424}$  &  $2.6_{-0.8}^{+1.4}$  &  2.7    \\
J2046-0421  &  $400_{-12}^{+235}$  &  $6.3_{-0.8}^{+6.2}$  &  16.7    \\
J2048-1616  &  $518_{-12}^{+435}$  &  $1.0_{-0.1}^{+1.2}$  &  2.8    \\
J2055+3630  &  $129_{-12}^{+376}$  &  $9.6_{-1.7}^{+2.5}$  &  9.5    \\
J2113+2754  &  $400_{-12}^{+188}$  &  $3.4_{-0.9}^{+1.8}$  &  7.3    \\
J2157+4017  &  $388_{-12}^{+471}$  &  $6.4_{-1.4}^{+2.2}$  &  7.0    \\
J2225+6535  &  $812_{-94}^{+153}$  &  $0.8_{-0.3}^{+0.5}$  &  1.1    \\
J2248-0101  &  $412_{-71}^{+235}$  &  $19.5_{-13.0}^{+-2.1}$  &  11.5    \\
J2305+3100  &  $529_{-165}^{+188}$  &  $5.9_{-0.7}^{+6.6}$  &  8.6    \\
J2346-0609  &  $729_{-59}^{+129}$  &  $5.3_{-0.6}^{+7.6}$  &  13.7    \\
\hline
\end{tabular}
\end{center}
\end{table}

\section{Posterior velocity distribution}
The posterior velocity distribution is useful for the forward and backward orbit integration as well as to estimate the effects of the source motion on its timing properties \citep{Shklovskii, 1994ApJ...421L..15C}. This is more important for millisecond radio pulsars because of their small magnetic fields. For one of the fastest radio pulsar in our sample (J1509+5531) the Shklovskii correction is $\Delta \dot P \approx 4\times 10^{-17}$ while its period derivative is $\dot P\approx 5\times 10^{-15}$. 

The posterior distribution is composed -- in accordance to the Bayesian theorem -- of a likelihood function and of a prior. The latter one is the optimal velocity distribution derived for the whole sample of young NSs. A use of prior is essential in the case of isolated radio pulsars since it supplies information about the missing radial velocity.

The likelihood function is the conditional probability to measure parallax $\varpi'$ and proper motion $\mu_\alpha'$, $\mu_\delta'$ given distance $D$, absolute value of velocity $v$, and velocity vector orientation angles $\xi_1, \xi_2$. These angles are the azimuth in the plane of sky ($0 \leq \xi_2 \leq 2\pi$ ) and the angle between line of sight and velocity vector ($0 \leq \xi_1 \leq \pi$). The measured values are considered to be independent, therefore the likelihood is a multiplication of independent probabilities:
$$
p(\varpi', \mu_{\alpha *}', \mu_\delta'| D, v, \xi_1, \xi_2) \propto g_D (\varpi' | D) \hspace{2.8cm}
$$
\begin{equation}
\hspace{2.8cm} \times g_\mu (\mu_{\alpha*}' | D, v, \xi_1, \xi_2)  g_\mu (\mu_\delta' | D, v, \xi_1, \xi_2)
\label{e:likelihood}
\end{equation}
where $g_D$ and $g_\mu$ are the normal distributions with zero mean  and standard deviations $\sigma_\varpi$, $\sigma_\alpha$ and $\sigma_\delta$ which correspond to observational uncertainties for parallax and proper motion respectively. These functions are written explicitly in \cite{verbunt2017}, see also a discussion about $g_D$ in \cite{2015PASP..127..994B} and \cite{igoshev2016}. 

\begin{table}
\caption{The numerical values for constants used in the analysis.}\label{t:num_val}
\begin{center}
\begin{tabular}{ccc}
\hline
\multicolumn{3}{c}{Galaxy rotation and local standard of the rest$^{1,2}$} \\
$R_\odot=8.5$~kpc    & $v_\odot = 220$~km~s$^{-1}$ & $h_\mathrm{OB} = 0.05$~kpc\\
$U = 10.0$~km~s$^{-1}$ & $V = 5.3$~km~s$^{-1}$ & $W=7.2$~km~s$^{-1}$ \\
\hline
\multicolumn{3}{c}{Prior velocity distribution$^3$} \\
$w = 0.42$ & $\sigma_1 = 75$ km s$^{-1}$ & $\sigma_2 = 316$ km s$^{-1}$  \\
\hline
\multicolumn{3}{c}{Pulsar braking$^4$} \\
$\kappa_0 = 1$ & $\kappa_1 = 1.4$ & $\kappa_2 = 1$ \\
\multicolumn{3}{c}{$\beta = 3\times 10^{-40}$ G s$^{-2}$} \\ 
\hline
\end{tabular}
\end{center}

1. \cite{DehnenBinney1998} 2. \cite{reed2000} 3. \cite{verbunt2017} 4. \cite{2014MNRAS.441.1879P}
\end{table}

The prior is a multiplication of functions describing the distance and velocity distribution (sum of two Maxwellians):
$$
p(D, v, \xi_1, \xi_2) = 2 f_D (D) \sin\xi_1  \sqrt{\frac{2}{\pi}}\left[ \frac{wv^2}{\sigma_1} \exp\left(-\frac{v^2}{2\sigma_1^2}\right)\right. \hspace{1.2cm}
$$
\begin{equation}
\hspace{3.2cm} \left.+\frac{(1-w)v^2}{\sigma_2} \exp\left(-\frac{v^2}{2\sigma_2^2}\right)\right]\Theta (zv_z)
\label{e:prior}
\end{equation}
The form of the spatial density $f_D(D)$ for radio pulsars was introduced in \cite{verbiest2012}. The theta function $\Theta (zv_z)$ of vertical height $z$ and vertical velocity $v_z$ implements the semi-isotropy condition: young pulsars move away from the Galactic plane. If older pulsars are considered, the theta function should be removed from the joint probability.
Values for $w, \sigma_1$ and $\sigma_2$ are summarized in Table~\ref{t:num_val}.

The joint probability $P_\mathrm{sim} (v, D, \xi_1, \xi_2, \varpi', \mu_{\alpha * }', \mu_{\delta}')$ is a multiplication of eq. (\ref{e:likelihood}) and (\ref{e:prior}) which is equal to eq. (28) in \cite{verbunt2017}.
Here we integrate the velocity orientation angles out:
\begin{equation}
P (v, D | \varpi', \mu_{\alpha * }', \mu_\delta') = \frac{\iint  P_\mathrm{sim}  d\xi_1 d\xi_2}{\iiiint P_\mathrm{sim} d\xi_1 d\xi_2 dD dv }
\label{e:poster_vD}
\end{equation}
The much simpler analytic eq. (19) from \cite{verbunt2017} written for the isotropic velocity distribution cannot be used because it depends on the velocity component in each direction and does not allow us to estimate the speed. 
The details of integration are presented in Appendix~\ref{a:first}. 

To get the posterior distribution for the absolute value of velocity, an additional integral is computed:
\begin{equation}
P(v) = \frac{\int P(v, D) dD}{\iint  P(v,D)dD dv} 
\end{equation}
This integral is easy to estimate based on previous calculations, simply adding up all posteriors values $P(v_i, D_j)$ for fixed velocity~$v_i$. 

\subsection{Results}
\begin{figure*}
\vspace*{0.4cm}
\begin{minipage}{0.48\linewidth}
\centerline{\includegraphics[width=1\columnwidth]{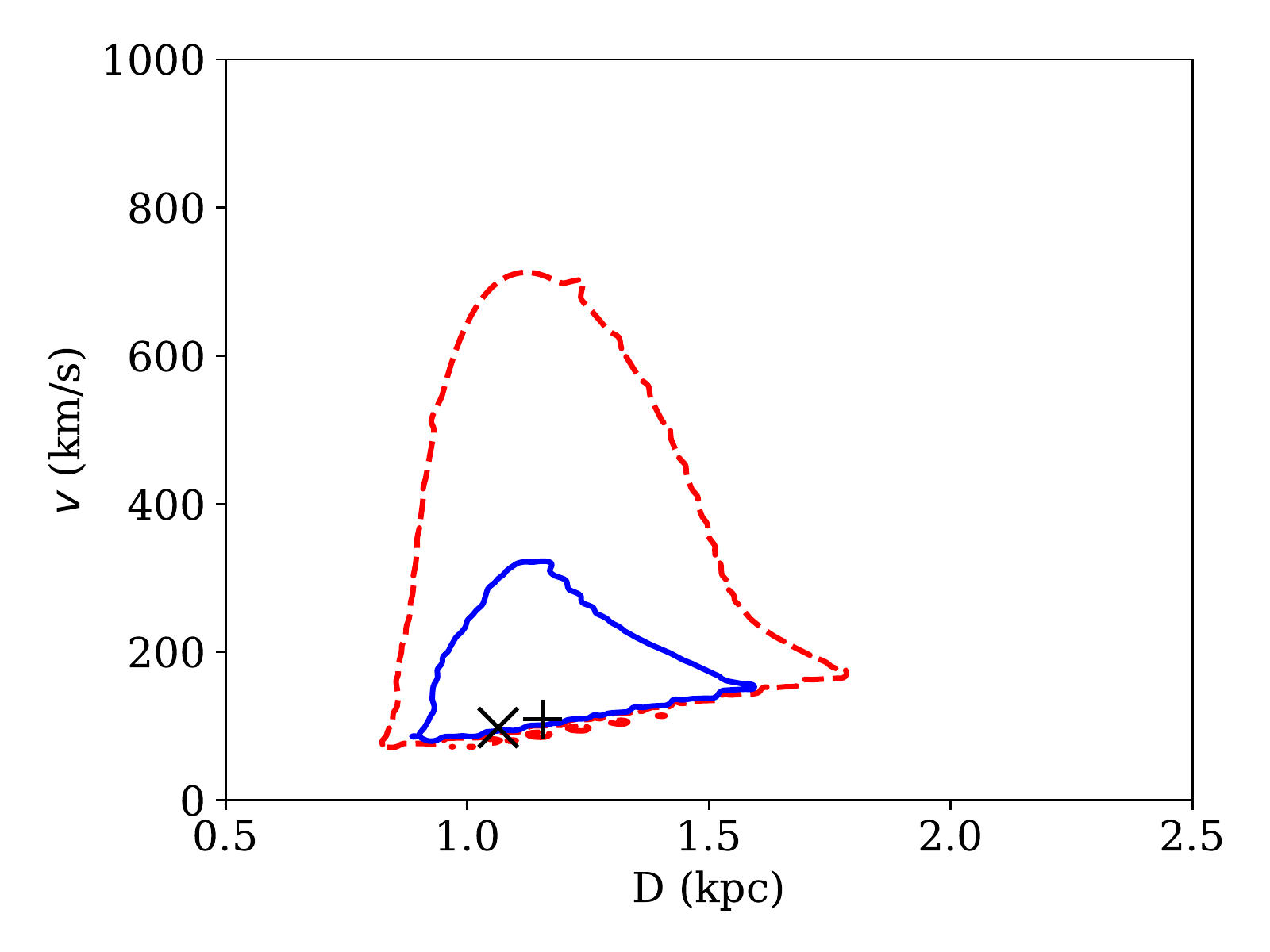}}
\end{minipage}
\begin{minipage}{0.48\linewidth}
\centerline{\includegraphics[width=1\columnwidth]{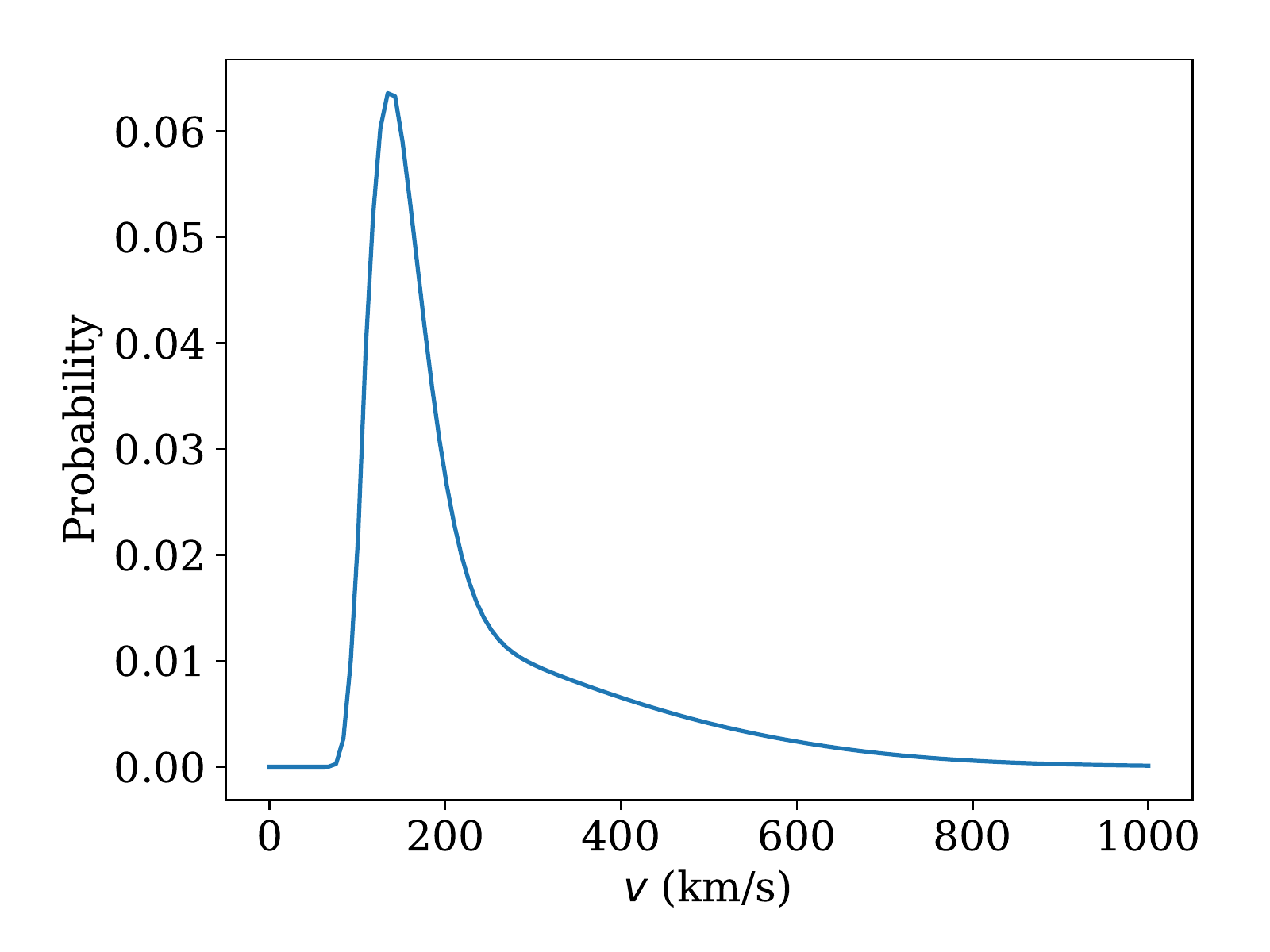}}
\end{minipage}
\caption{The posterior distribution $P(D, v)$ (left panel) and $P(v)$ (right panel) for PSR J0332+5434. On left panel the cross 
corresponds to the nominal value of the  distance $D' = 1/\varpi'$ and transversal velocity $v_t'=\kappa \mu_t' /\varpi'$ while "+" sign shows the most probable value for distance and velocity. Red dashed and blue solid contours correspond to 68\% and 95\% of the integrated probability respectively.
\label{f:posterior_vD}}
\end{figure*}

An example of the posterior velocity distribution for PSR J0332+5434 based on its parallax and proper motion measurements from \cite{2002ApJ...571..906B} is presented in Figure \ref{f:posterior_vD}.
The function peaks close to the nominal value of the distance $D' = 1/\varpi'$ and transversal velocity $v'=\kappa \mu_t'/\varpi'$  where $\kappa = 4.74$~km~s$^{-1}$~yr~kpc$^{-1}$ is a unit conversion coefficient. It follows a line $v = \kappa \mu_t' D$\footnote{Further we skip $\kappa$ to simplify equations. The correct values can be reproduced if it is assumed $\mu_b' = \kappa \mu_{b,\mathrm{meas}}$ and $\sigma_b = \kappa \sigma_{b,\mathrm{meas}}$ where $\mu_{b,\mathrm{meas}}$ and $\sigma_{b,\mathrm{meas}}$ are the measured value. }. The probability density has a long tail in the direction of large velocities because the radial component is not measurable and is drawn from the sum of two Maxwellians. When the velocity is used to estimate the kinematic ages, the tail contributes to the shortest age estimates.
Values of the velocity computed with resolution 12~km~s$^{-1}$ are summarized in Table~\ref{t:res} together with their 68 per cent credible intervals.   

\section{Posterior kinematic age distribution}
The kinematic age of a young radio pulsar  ($\tau < 20$~Myr, see Section~\ref{s:gp_inf} for discussion) at Galactic latitude $b$ is defined as:
\begin{equation}
t_\mathrm{kin}(D, z_0, v_b, v_r) = \frac{D\sin b - z_0}{v_b\cos b + v_r \sin b}
\label{e:tkin}
\end{equation}
where the actual velocities $v_b$ and $v_r$ are in the Galactic latitude and the radial direction respectively corrected for the motion of the local standard of rest. It is assumed that the pulsar was born at a distance $z_0$ from Galactic plane. This estimate works only for pulsars with noticeable motion directed away from the Galactic plane. The immediate consequence of eq.~(\ref{e:tkin}) is that an unknown radial velocity is getting more important for older pulsars which are far away. 

The eq. (\ref{e:tkin}) depends on actual distance, radial and latitudinal velocities and birth height. These are considered to be random unknown values which are specified by setting the priors and likelihood functions. As soon as posteriors for each of these variables are constructed, we can draw a sample from each posterior and estimate the age for each individual element from the sample using eq. (\ref{e:tkin}). We follow this path performing as many steps analytically as it is possible.

The deterministic relation between age, distance and velocity makes us to write $p(t|D,z_0, v_b, v_r)$ in form of a delta function. To avoid dealing with delta function we can also introduce the normal distribution with standard deviation $\sigma_t$ which tends to zero:
\begin{equation}
p(t | D, z_0, v_b, v_r)dt = \frac{1}{\sqrt{2\pi}\sigma_t} \exp\left[ -\frac{1}{2} \frac{(t - t_\mathrm{kin}(D, z_0, v_b, v_r))^2}{\sigma_t^2}\right] dt
\end{equation}
The joint probability is written as:
$$
P(\varpi', \mu_b', D, z_0, v_b, v_r,t) \propto g_D (\varpi'|D) f_D(D) g_\mu (\mu_b' | \mu_b) p(t|D,z_0,v_b,v_r)
$$
\begin{equation}
\hspace{0.25cm}\times f_z(z_0)\left[w G(v_b | \sigma_1) G(v_r | \sigma_1) + (1-w)G(v_b | \sigma_2) G(v_r | \sigma_2)\right] 
\label{e:joint}
\end{equation}
where $G(x | \sigma_x)$ is zero-centered normal distribution in form:
\begin{equation}
G(x|\sigma_x)dx = \frac{1}{\sqrt{2\pi}\sigma_x} \exp \left[ -\frac{1}{2} \frac{x^2}{\sigma_x^2}\right]dx
\end{equation}
The prior distribution for birth heights is exponential:
\begin{equation}
f_z(z_0) = \frac{1}{h_\mathrm{OB}} \exp\left[-\frac{|z|}{h_\mathrm{OB}}\right]
\end{equation}
The value of $h_\mathrm{OB}$ is given in Table~\ref{t:num_val}. The prior in this form includes also disadvantageous birth positions such as birth location below the Galactic plane for pulsar which is observed above the Galactic plane now.
The prior $f_z(z_0)$ complements $f_D (D)$ since we are interested in relating NS with their progenitors which have different Galactic scale height than currently observed pulsar ensemble. After the joint probability eq.~(\ref{e:joint}) is written, we start deriving the posterior distribution.


\subsection{Kinematic age with accurate latitudinal velocity, birth height and distance}
\begin{figure*}
\begin{minipage}{0.48\linewidth}
\centerline{\includegraphics[width=1\columnwidth]{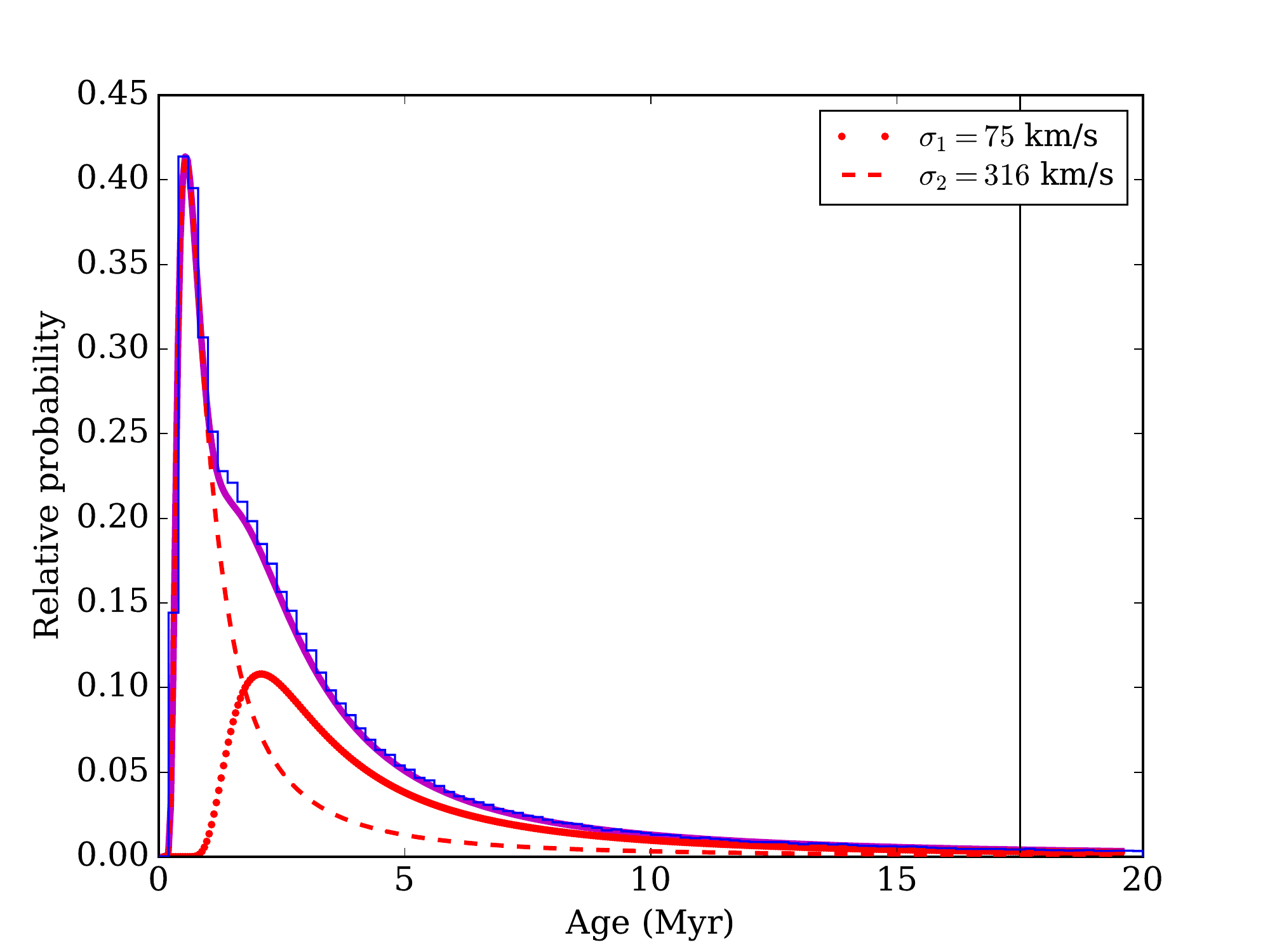}}
\end{minipage}
\begin{minipage}{0.48\linewidth}
\centerline{\includegraphics[width=1\columnwidth]{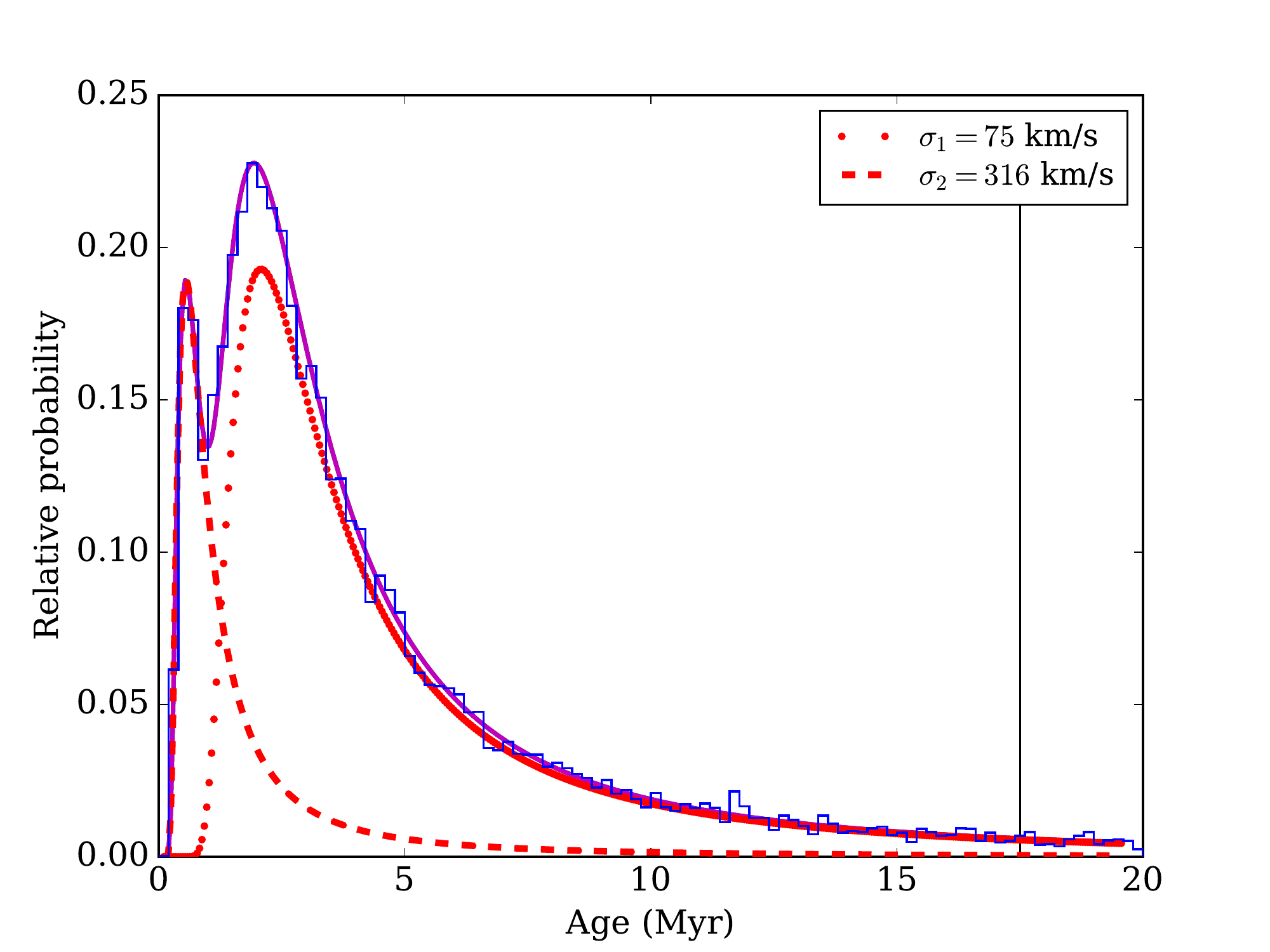}}
\end{minipage}
\caption{The posterior distribution of kinematic ages $P(t| D, v_b)$ (left panel) and $P(t | D, \mu_b')$ (right panel) for B0950+08. The black line 
corresponds to the spin-down age. The histograms show the results of the Monte Carlo simulations. Red lines shows the contribution of low and high-velocity components of the radial velocity prior to the posterior. 
\label{f:posterior_tkin}}
\end{figure*}

The first step is a pure mathematical exercise which is added here to show the essential role of unknown radial velocity in the kinematic age estimate. 
The simplest kinematic age estimate is based on nominal values of distance, latitudinal velocity and the radial velocity distribution. To get this estimate we  integrate the eq.~(\ref{e:joint}) over the unknown radial velocities:
\begin{equation}
p(\varpi', \mu_b', D, z_0, v_b, t) \propto \int_{-\infty}^\infty P(\varpi', \mu_b', D, z_0, v_b, v_r,t) dv_r
\label{e:zero_step}
\end{equation}
The integral is split into a sum of two integrals: each of them corresponds to one mode of the velocity distribution.
Inside both integrals the terms $g_\mu(\mu_b'|\mu_b) G(v_b | \sigma)g_\varpi (\varpi'|D) f_D(D) f_z(z_0)$ do not depend on radial velocity and are moved outside of the integral. The remaining part is:
\begin{equation}
p(t|D, z_0, v_b) \propto \int_{-\infty}^\infty G(v_r | \sigma) p(t | D, z_0, v_r, v_b) dv_r
\label{e:first_step}
\end{equation}
This integral can be computed analytically:
$$
p_1(t|D, z_0, v_b) \propto \frac{w}{2\pi \sigma_1 \sigma_t} \int_{-\infty}^\infty \exp\left[-\frac{v_r^2}{2\sigma^2} - \frac{(t - t_\mathrm{kin}(D,v_b,v_r))^2}{2\sigma_t^2}\right]dv_r
$$
\begin{equation}
=\frac{w(D-z_0 / \sin b )}{\sqrt{2\pi}\sigma_1 t^2} \exp\left[-\frac{(D/t - z_0 \csc b /t - v_b \cot b)^2}{2\sigma_1^2}\right]
\end{equation}
Here we assume that $\sigma_t \to 0$ and the normal distribution which includes $\sigma_t$ is properly normalized. 
The total conditional probability for the case $z_0=0$ is a sum of two modes:
$$
p(t|D, v_b)dt =\frac{w}{\sqrt{2\pi}\sigma_1} \frac{D}{t^2} \exp\left[-\frac{(D/t - v_b \cot b)^2}{2\sigma_1^2}\right] dt\hspace{1.2cm}
$$
\begin{equation}
\hspace{2.4cm}+\frac{1-w}{\sqrt{2\pi}\sigma_2} \frac{D}{t^2} \exp\left[-\frac{(D/t - v_b \cot b)^2}{2\sigma_2^2}\right] dt
\label{e:cnd_vb_d}
\end{equation}
An example of this conditional probability is plotted in Figure~\ref{f:posterior_tkin} (left panel) for PSR B0950+08 ($b=43\fdg 70$) with fixed nominal distance $D' = 1/\varpi'=0.26$~kpc and $v_b' = \kappa \mu_b' /  \varpi' = 16.1$~km~s$^{-1}$. The value of $v_b'$ must be corrected for the Solar motion in the Galaxy and the local standard of rest for pulsar. This correction gives $v_b'' = 30.3$~km~s$^{-1}$.

The final distribution eq.~(\ref{e:cnd_vb_d}) in principle represents the result of  Monte Carlo simulation where $v_r$ is drawn from a sum of two Maxwellians with fixed $w, \sigma_1$ and $\sigma_2$, and the age is computed according to eq.~(\ref{e:tkin}) with fixed $D, b$ and $v_b$, see histogram in Figure~\ref{f:posterior_tkin} (left panel).

\subsection{Kinematic age with accurate distance and birth height}
The latitudinal velocity and its uncertainty constrain the radial component only if the prior for the velocity distribution is bimodal i.e. all velocity components can be chosen from either low or high velocity component of a sum of two Maxwellian distributions.
If prior is a single Maxwellian distribution, the value of $v_b$ sets no constraints on value of $v_r$. 
The optimal model for the velocity distribution of young radio pulsars in the article by \cite{verbunt2017} contains two separate modes i.e. all three components of the velocity have to belong either to the low or high-velocity Maxwellian. This choice favors the physical explanation with two separate formation mechanisms (e.g. core collapse and e-capture supernova explosion) or single formation mechanism with two channels (whether $l=1$ or $l=2$ dominates in a shock instability e.g. \citealt{2005ASPC..332..363J})  and disfavors a wide velocity distribution. 

The eq. (\ref{e:zero_step}) is integrated further over uncertain latitudinal velocity $v_b$:
\begin{equation}
P(\varpi', \mu_b', D, z_0, t) \propto \int_{-\infty}^\infty  P(\varpi', \mu_b', D,z_0,v_b, t) dv_b
\end{equation}
The terms $g_\varpi (\varpi'|D) f_D(D)$  and $f_z(z_0)$ do not depend on $v_b$, so we move them out of integral. If we drop these terms for a moment, we can write the conditional probability to measure the kinematic age for a single mode of the velocity prior for a fixed actual distance as:
$$
p_1(t|D,\mu_b') \propto \frac{1}{\sqrt{8\pi^3}\sigma_1^2 \sigma_b} \frac{D}{t^2} \int_{-\infty}^\infty \exp\left[-\frac{(D/t - v_b \cot b)^2}{2\sigma_1^2}\right] \hspace{1cm}
$$
\begin{equation}
\hspace{1.5cm}\times \exp\left[ -\frac{v_b + D(\mu_{Gb} - \mu_b')^2}{2D^2\sigma_b^2} \right] \exp\left[ -\frac{v_b^2}{2\sigma_1^2}\right] dv_b
\label{e:cnd_t_d}
\end{equation}
where $\sigma_b$ is an uncertainty of $\mu_b'$ measurement and $\mu_{Gb}$ is correction for the Galactic rotation and peculiar velocity of the Sun.
This form allows a combination of $(v_r, v_b)$ where both terms are drawn either from a Maxwellian with the standard deviation $\sigma_1$ or  $\sigma_2$. 
The integral in eq. (\ref{e:cnd_t_d}) is computed analytically. To do so, we introduce auxiliary variables:
\begin{equation}
A_1 = \frac{1}{2\sigma_1^2} + \frac{\cot^2 b}{2\sigma_1^2} + \frac{1}{2\sigma_b^2 D^2}
\label{e:a1}
\end{equation}
\begin{equation}
B_1 = - \frac{(D-z_0/\sin b)\cot b}{\sigma_1^2 t} + \frac{\mu_{Gb} - \mu_b'}{D\sigma_b^2}
\label{e:b1}
\end{equation}
\begin{equation}
C_1 = \frac{(D - z_0/\sin b)^2}{2t^2 \sigma_1^2} + \frac{(\mu_{Gb} - \mu_b')^2}{2\sigma_b^2}
\label{e:c1}
\end{equation}
In this case the result of the integration is written as:
$$
p(t|D,\mu_b')dt \propto \frac{w}{\sqrt{8}\pi\sigma_1^2 \sigma_b} \frac{D - z_0 / \sin b}{t^2\sqrt{A_1}} \exp \left[ \frac{B_1^2}{4A_1} - C_1 \right] dt \hspace{1.2cm}
$$
\begin{equation}
\hspace{1.7cm}+ \frac{1-w}{\sqrt{8}\pi\sigma_2^2 \sigma_b} \frac{D-z_0/\sin b}{t^2\sqrt A_2} \exp \left[ \frac{B_2^2}{4A_2} - C_2 \right]dt
\label{e:cnd_v_d_final}
\end{equation}
where auxiliary variables with subscripts 1 and 2 stand for ones computed with $\sigma_1$ and $\sigma_2$.
The result of the calculation according to this equation for PSR B0950+08 are shown in Figure~\ref{f:posterior_tkin} (right panel). The contribution of the second mode has strongly increased comparatively to the previous case, see left panel of the same Figure. The reason for this is that latitudinal velocity $v_b''=30.3$~km~s$^{-1}$ is small, so the velocity of a pulsar is more probable to be drawn from the low-velocity than from the high-velocity component of the Maxwellian.

We test the results of the integration by performing a Monte Carlo simulations. We draw pairs $v_r, v_b$ from a Gaussian with $\sigma_1$ in $w$ cases and from Gaussian with $\sigma_2$ in $1-w$ cases. After this we fix the distance at its nominal value $D' = 1/\varpi' = 0.26$~kpc and select pairs $v_r, v_b$ 
according to the normal distribution:
\begin{equation}
f(\mu_{b,\mathrm{gen}}) = \exp\left[-\frac{1}{2}\frac{(\mu_{b,\mathrm{gen} - \mu_b'})^2}{\sigma_b^2}\right]
\end{equation}
where the proper motion $\mu_{b,\mathrm{gen}} = v_b / (\kappa D$). It means that we preferably leave in the sample $(v_r, v_b)$ pairs which give the proper motion in the latitudinal direction close to measured one. 
For all such pairs of $v_r, v_b$ we compute the kinematic age using eq. (\ref{e:tkin}). In Figure~\ref{f:posterior_tkin} (right panel) we show  results of Monte Carlo simulation for B0950+08. The analytical probability density follows the result of the Monte Carlo simulations with high precision.


\subsection{Complete description}

\begin{figure}
\centerline{\includegraphics[width=1\columnwidth]{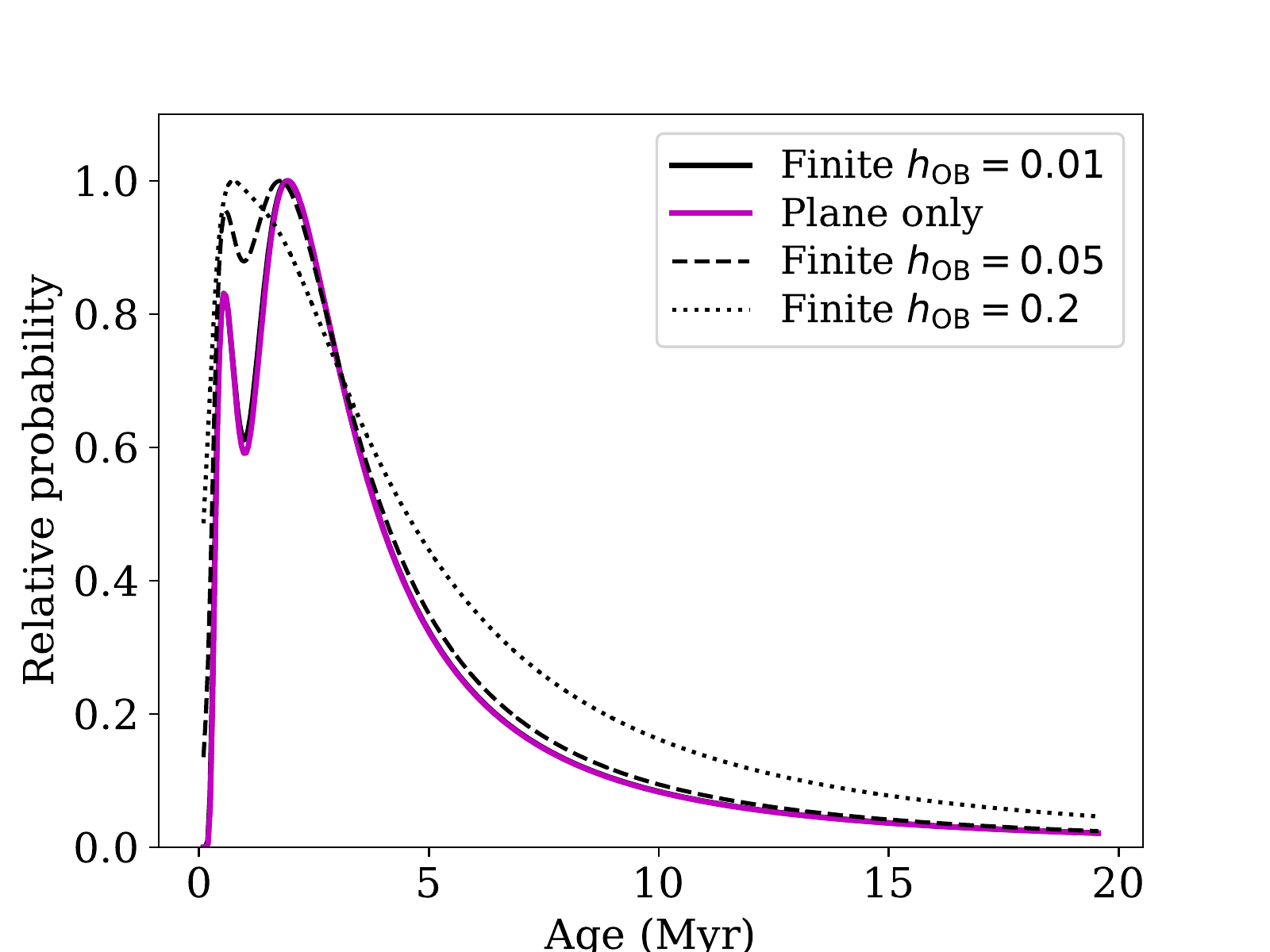}}
\caption{The posterior distribution of kinematic ages $P(t| \varpi', \mu_b')$ for B0950+08. 
\label{f:posterior_tkin_final}}
\end{figure}

The rigorous derivation of the posterior distribution for the case of $h_\mathrm{OB} \neq 0$ (progenitors could be born above and below the Galactic plane) are summarized in Appendix~\ref{a:thin}. The marginal over all variables posterior kinematic age is:
\begin{equation}
P(t | \varpi', \mu_b')dt \propto \int_0^{D_\mathrm{max}} p(t | D, \mu_b') g_\varpi (\varpi' | D) f_D(D) dD
\end{equation}
where $D_\mathrm{max}=10$~kpc. This integral is computed numerically using the Gauss quadrature method with 64 nodes.
For pulsars with parallaxes measured by the interferometric technique, the contribution of the distance uncertainty is quite small, see an example in Figure~\ref{f:posterior_tkin_final} which is quite similar to Figure~\ref{f:posterior_tkin}. The consideration of $z_0 \neq 0$ changes the posterior distribution very little if we use realistic value  $h_\mathrm{OB}=0.05$~kpc \citep{reed2000}.

We compile estimates of the kinematic ages and its credible intervals in the Table~\ref{t:res}.
The values presented in the Table~\ref{t:res} are in agreement within error bars with ones published in \cite{Noutsos2013}. There is a a single exception of PSR J1932+1059 which is one order of magnitude younger in our article. This happens because we took into account only the solution with the smallest age. We assume that pulsar did not have enough time to oscillate in the Galactic gravitational potential, while \cite{Noutsos2013} assumed a solution with larger age.

\begin{figure}
\centerline{\includegraphics[width=1\columnwidth]{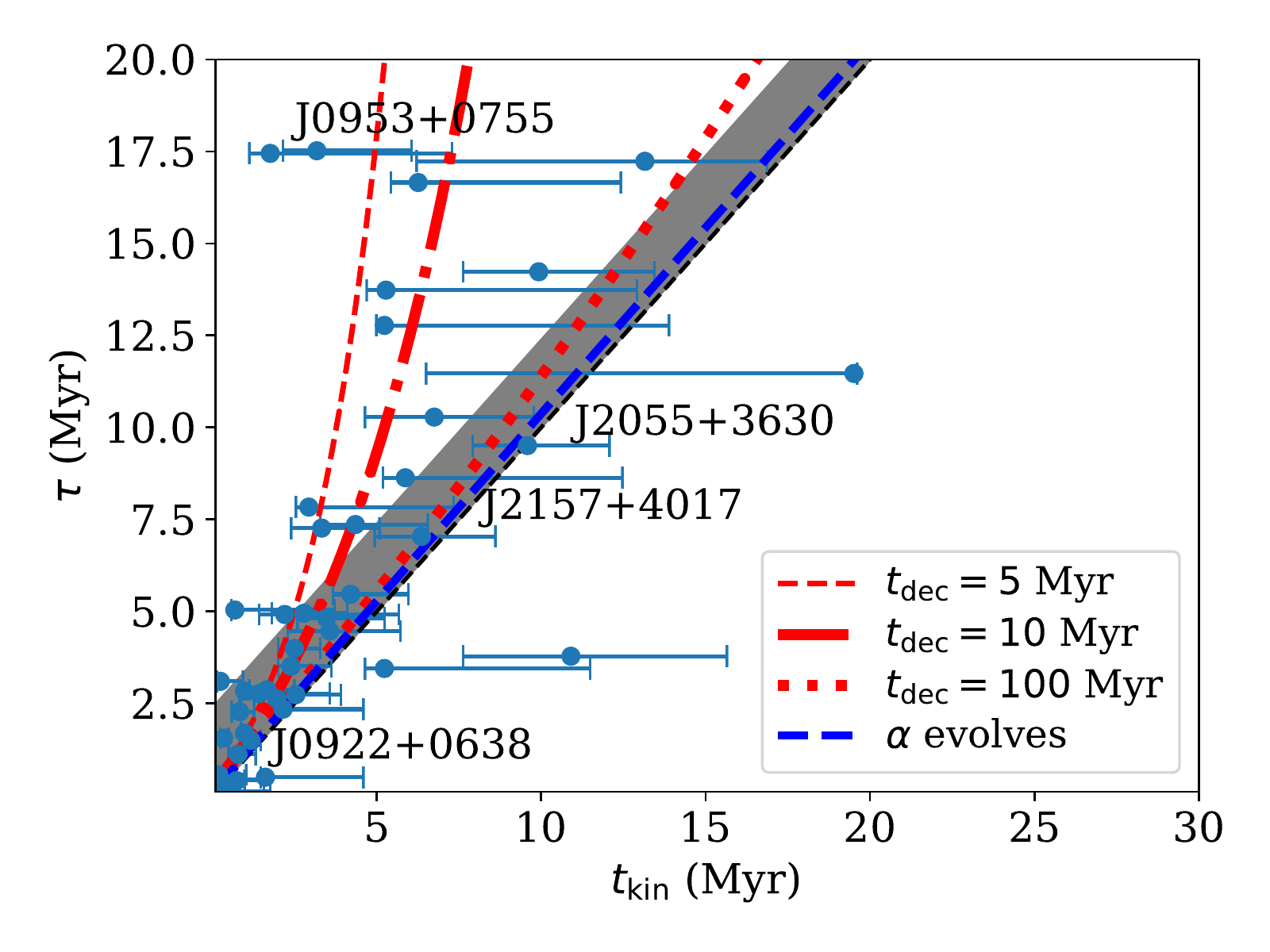}}
\caption{Relation between the spin-down and kinematic ages. The error bars correspond to $68\%$ credible interval. Grey region indicates a typical effect of large initial periods, red lines show the evolution of the spin down ages in a case of magnetic field decay with different typical timescales. Dashed blue line shows possible effect of the obliquity angle evolution according to the article by \protect\cite{2014MNRAS.441.1879P}. 
\label{f:t_tau}}
\end{figure}

We plot  spin-down and kinematic ages with its 68\% credible interval in Figure~\ref{f:t_tau}. It is not surprising that pulsars with large $\tau$ have larger uncertainty in the kinematic age. Such pulsars typically traveled far away from the Galactic plane which translates into large $b$, so the contribution of unknown radial velocity start playing significant role. 


\subsection{Influence of the Galactic gravitational potential}
\label{s:gp_inf}
The estimate of the kinematic age in form of eq. (\ref{e:tkin}) is  the first term of the Taylor series:
\begin{equation}
t_\mathrm{kin} = \frac{z}{v_z} = \frac{z}{v_{z, 0} + \dot v_{z, 0} \Delta t + ...} \approx \frac{z}{v_{z,0}(1 + \dot v_z \Delta t/v_{z,0})} 
\end{equation}
The second term becomes crucial when:
\begin{equation}
\Delta t \approx \frac{v_{z,0}}{|\dot v_z|} = v_{z,0} \left|\frac{\partial \Phi}{\partial z} \right|^{-1}
\end{equation}
A typical value for the gravitational force in z-direction in the solar vicinity at distance $z = 100$ pc is 0.9 km~s$^{-1}$~Myr$^{-1}$ which means that the kinematic age estimate  eq. (\ref{e:tkin}) is applicable for ages $t < 20$ Myr and velocities $v_z > 80$ km~s$^{-1}$ (possible correction less than $\approx 20$ per cent). In reality all slow objects in Table~\ref{t:res} have small spin-down ages ($\tau < 3$~Myr) except for J0953+0755.

\begin{figure*}
\begin{minipage}{0.48\linewidth}
\centerline{\includegraphics[width=1\columnwidth]{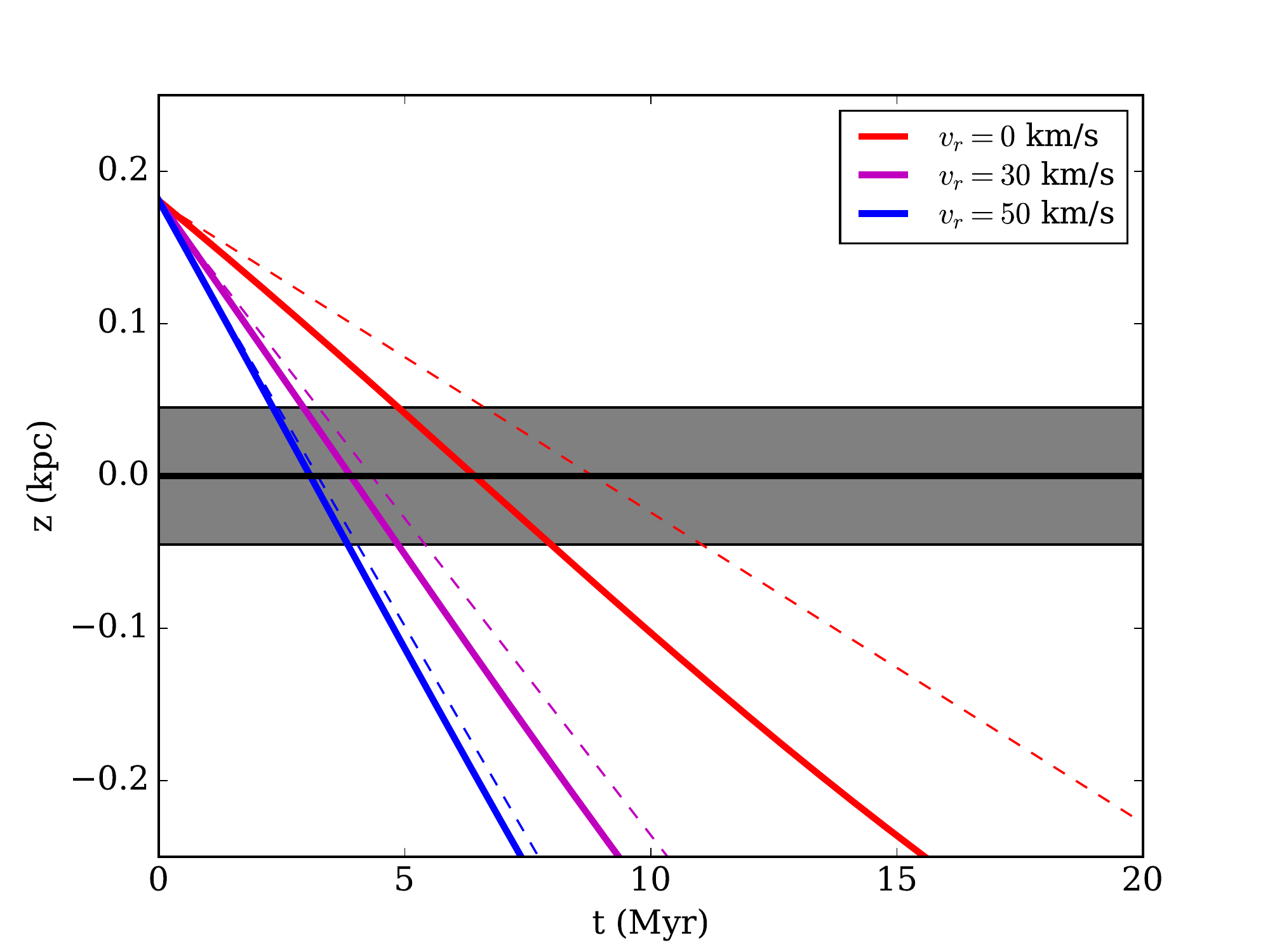}}
\end{minipage}
\begin{minipage}{0.48\linewidth}
\centerline{\includegraphics[width=1\columnwidth]{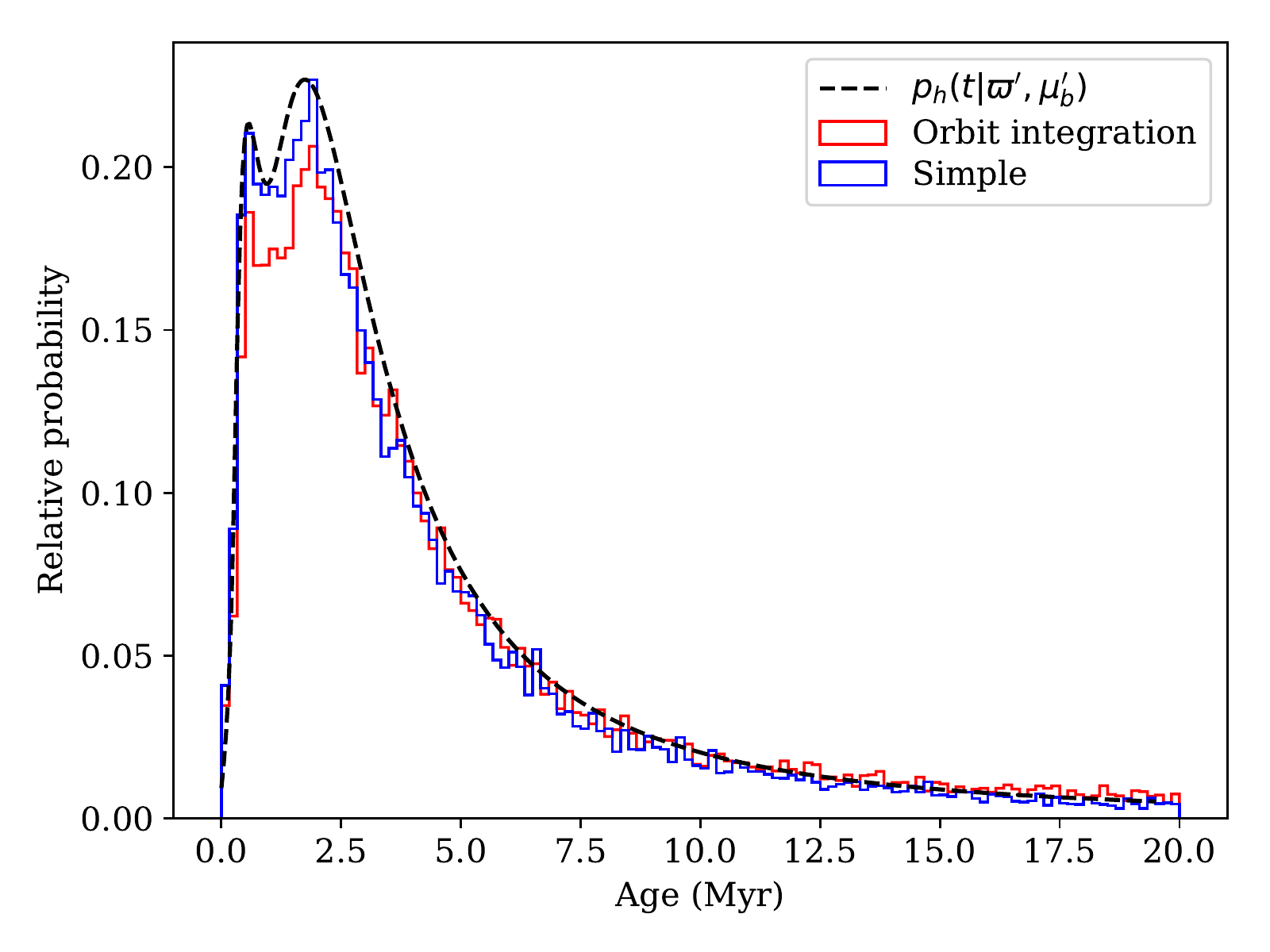}}
\end{minipage}
\caption{The motion of the PSR B0950+08 in the vertical direction in the Galactic gravitational potential (left panel). The color solid lines show the result of the numerical integration. Posterior for the kinematic age derived by backward integration in realistic Galactic gravitational potential (right panel).  
\label{f:gp}}
\end{figure*}

 The comparison between real physical trajectory and its simple estimate is shown in Figure~\ref{f:gp} for PSR B0950+08. For larger radial velocities ($v_r > 30$ km~s$^{-1}$) the difference is negligible. For small and negative radial velocities the difference is dramatic. It means that the estimate based on eq. (\ref{e:tkin}), in particular analytical eq. (\ref{e:cnd_v_d_final}) can be applied only to pulsars which show clear evidence of young ages: namely $\tau < 20$~Myr and they move away from the Galactic plane with noticeable speed. 

To  check carefully how appropriate is the age estimate in the case of PSR B0950+08, we perform the Markov Chain Monte Carlo (MCMC) simulations using the backward integration. 
We sample the probability density in form of eq.~(\ref{e:joint}) by means of the MCMC sampler \texttt{emcee} based on \cite{2010CAMCS...5...65G} algorithm. We  use 24 walkers and generate chain with length 5000 out of which the first 500 items are discarded to guarantee that chains fill the whole parametric space. Instead of the simple estimate for the kinematic age in form of eq.~(\ref{e:tkin}) we use the backward integration in the Galactic gravitational potential \texttt{MWPotential2014} from package \texttt{galpy} \citep{bovy}\footnote{http://github.com/jobovy/galpy}. 
Each orbit is integrated for 120~Myr with 2000 integration steps. 
The first moment of crossing $z_0$ is refined by means of linear interpolation and recorded. 
We check the convergence of the MCMC process by two independent tests: (1) we compute the integrated auto-correlation time which consists of 78 elements ($\approx 58$ independent samples) and (2) we compute the ages based on eq.~(\ref{e:tkin}) for all generated initial conditions and check that it closely follows the analytically derived probability density, see Figure~\ref{f:gp}.

Based on the MCMC simulations we  find the most probable age of B0950+08 to be 1.76~Myr (Figure~\ref{f:gp}). The credible interval which contains $68$ per cent of the probability density is [1.2,8.0]~Myr, the credible interval which contains $95$ per cent of the probability is [0.37, 17.0]~Myr. The simple analytical estimates agrees with this rigorous one taking into account uncertainty ranges.


\section{Magnetic field evolution}
The combination of kinematic and spin-down ages for a large number of pulsars allows us to test the magnetic field evolution which controls the spin-down age.
To do it in quantitative way, we assume the exponential magnetic field decay:
\begin{equation}
B(t) = B_0 \exp\left(-\frac{t}{t_\mathrm{dec}}\right)
\end{equation}
In this case the spin-down age evolves as:
\begin{equation}
\tau(t) = \left[\tau_0 + \frac{t_\mathrm{dec}}{2}\right] \exp\left(\frac{2t}{t_\mathrm{dec}}\right) - \frac{t_\mathrm{dec}}{2} 
\label{e:tau_t}
\end{equation}
where $\tau_0 = \beta P_0^2 / (2 B_0^2)$ is the initial spin-down age composed of the initial period $P_0$ and initial magnetic field $B_0$.

As soon as $t_\mathrm{dec} \gg t$ it means that $\tau(t)\approx t+\tau_0$. When $t > t_\mathrm{dec}$ we start seeing the exponential growth of the spin-down age. We plot the curves $\tau(t)$ for different magnetic field decay timescales in Figure~\ref{f:t_tau}. It is clear that the decay timescale cannot be less than ten Myr. The strongest restrictions on this value comes from PSR J2055+3630 and J2157+4017 for which $\tau \approx t$. The longer decay time $t_\mathrm{dec} > 100$ Myr is impossible to probe with this method.
In the range of interest $1-20$ Myr the neutron star crust is cold which means that the phonon resistivity in the crust plays no role. The essential contribution to crust resistivity originates from the crust impurity. The surface magnetic field could also be affected by magnetic field evolution in the NS core, see e.g. \cite{2015MNRAS.453..671G}. 

The initial period and magnetic field contribute to the spin-down age. 
In the case of $B_p = 10^{12}$~G the initial spin-down age translates to 0.26 Myr for $P_0 = 0.1$ s and to 2.37 Myr for $P_0 = 0.3$ s. It can cause a shift in $\tau$ relatively to $\tau=t$ line, see the gray region in Figure~\ref{f:t_tau}. 

Quantitative description can be derived if we invert the eq.~(\ref{e:tau_t}):
\begin{equation}
t(\tau, t_\mathrm{dec},\tau_0) = \frac{t_\mathrm{dec}}{2} \log \left[\frac{\tau-0.5 t_\mathrm{dec}}{\tau_0 + 0.5t_ \mathrm{dec}}\right]
\end{equation}
In principle, this function can be fitted to the data points at the Figure~\ref{f:t_tau} by means of the least square technique to estimate the magnetic field decay timescale. This approach is highly inefficient because it assumes the normal distribution for uncertainties in the kinematic age and $\tau_0 \ll \tau$ or alternatively the same $\tau_0$ for all objects.

Instead, we develop a maximum likelihood approach which makes use of complete joint probability density $p(\varpi', \mu_b', t)$ and estimates the distribution of $\tau_0$ based on earlier works. The derivations start from the joint probability:
\begin{equation}
p(\varpi', \mu_b', \tau, t, \tau_0 | t_\mathrm{dec}) = p(\varpi', \mu_b', t) p(\tau | t_\mathrm{dec}, \tau_0, t) p(\tau_0)
\label{e:joint_age}
\end{equation}
The relation $ p(\tau | t_\mathrm{dec}, \tau_0, t) $ is analytical and can be written as delta function or a normal distribution with the standard deviation which tends to zero:
\begin{equation}
 p(\tau | t_\mathrm{dec}, \tau_0, t)  = \frac{1}{\sqrt{2\pi}\sigma_t} \exp\left(-\frac{(t - t(\tau, t_\mathrm{dec},\tau_0))^2}{2\sigma_t^2}\right)
\end{equation}
The eq.~(\ref{e:joint_age}) is integrated two times: (1) over ages and (2) over initial spin-down ages:
\begin{equation}
p(\varpi', \mu_b', \tau | t_\mathrm{dec}) = \iint p(\varpi', \mu_b', \tau, t, \tau_0 | t_\mathrm{dec})dt d\tau_0
\end{equation}
The integral over ages is computed analytically which leads to:
\begin{equation}
p(\varpi', \mu_b', \tau | t_\mathrm{dec}) = \int p(\varpi', \mu_b', t (\tau, t_\mathrm{dec} \tau_0)) p(\tau_0) d\tau_0
\label{e:integ_tau}
\end{equation}
This integral is computed numerically using following prescription. First, we draw ten millions of $P_0$ and $B_0$ based on measurements from \cite{popov2012} and compute $\tau_0$. Second, we bin $\tau_0$ in bins of $0.4$~Myr and use a linear interpolation to create numerical $p(\tau_0)$. Third, for each pulsar, the integral eq.~(\ref{e:integ_tau}) is computed from $\tau_0 = 0$ till $\tau_0 = \tau'$. 

The integral  eq.~(\ref{e:integ_tau}) is a likelihood for parameter $t_\mathrm{dec}$ of individual pulsar. The total log-likelihood is a sum of log-likelihoods for all pulsars:
\begin{equation}
L(t_\mathrm{dec}) = \sum_{i=1}^N \log\left[p(\varpi_i', \mu_{b,i}', \tau_i | t_\mathrm{dec})\right]
\end{equation}
There is one important caveat: the $p(\tau_0)$ is not known with such a great precision. Therefore, the confidence limits estimated this way are rather indicative than precise. 

The maximum likelihood analysis is tested on synthetic samples prepared in following manner: real ages are drawn from the uniform distribution (0, 20)~Myr. For all objects, we assume the same $t_\mathrm{dec}$ ranging from 3~Myr to 18~Myr in different samples. The initial spin-down age is drawn the same manner as in analysis. After $\tau$ is computed for each object in the synthetic sample, we assign the normal distribution for real ages centered on generated value with the standard deviation which grows linearly with the real age.  By testing the method, it become clear that (1) method estimates the $t_\mathrm{dec}$ precisely, (2) if proper treatment of the $\tau_0$ is not included in the likelihood (i.e. it is assumed that $\tau_0 = 0$), the method underestimates the $t_\mathrm{dec}$ up to 2-3 times and provides too restrictive confidence limits. 

\begin{figure*}
\begin{minipage}{0.48\linewidth}
\centerline{\includegraphics[width=1\columnwidth]{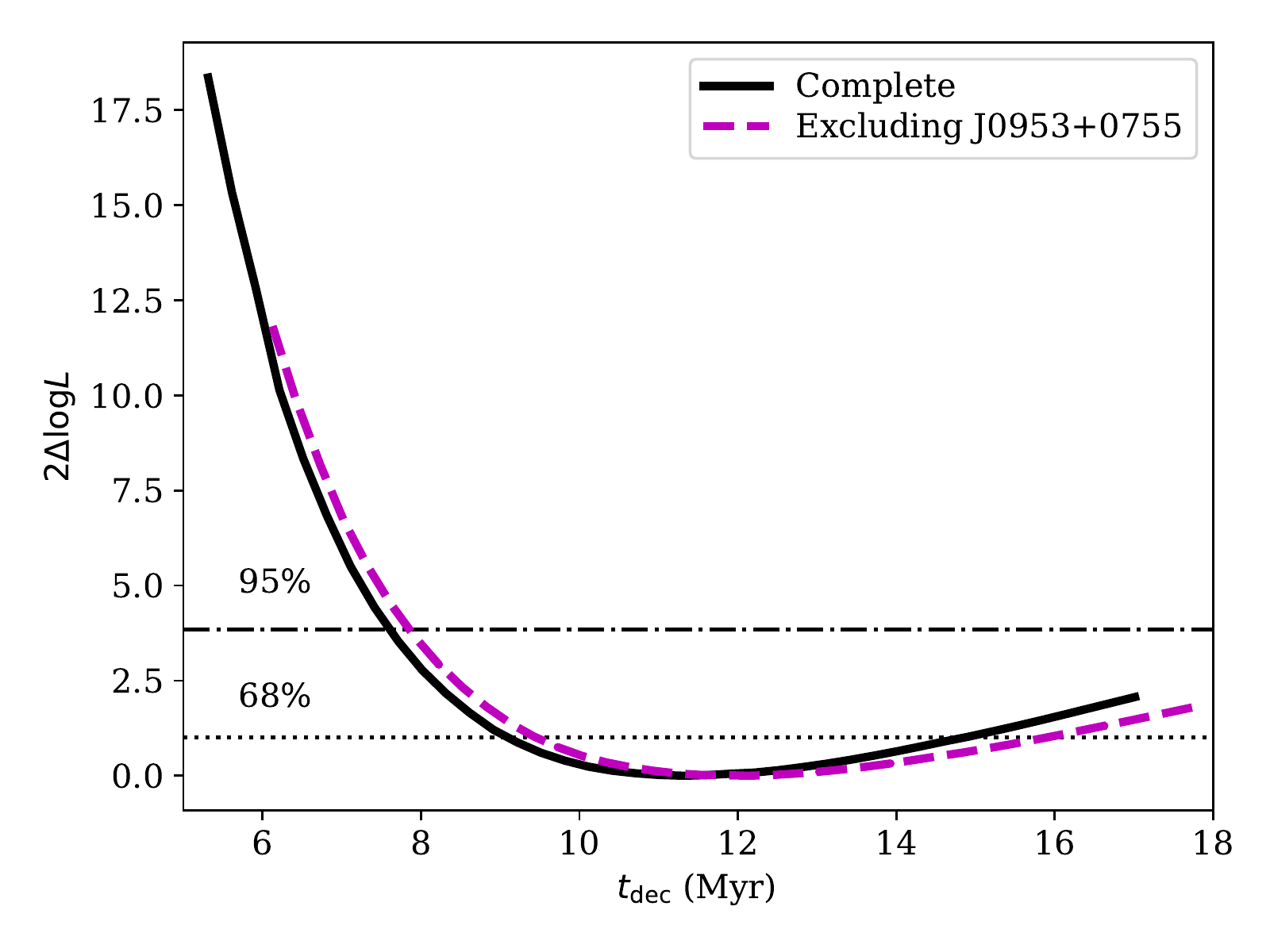}}
\end{minipage}
\begin{minipage}{0.48\linewidth}
\centerline{\includegraphics[width=1\columnwidth]{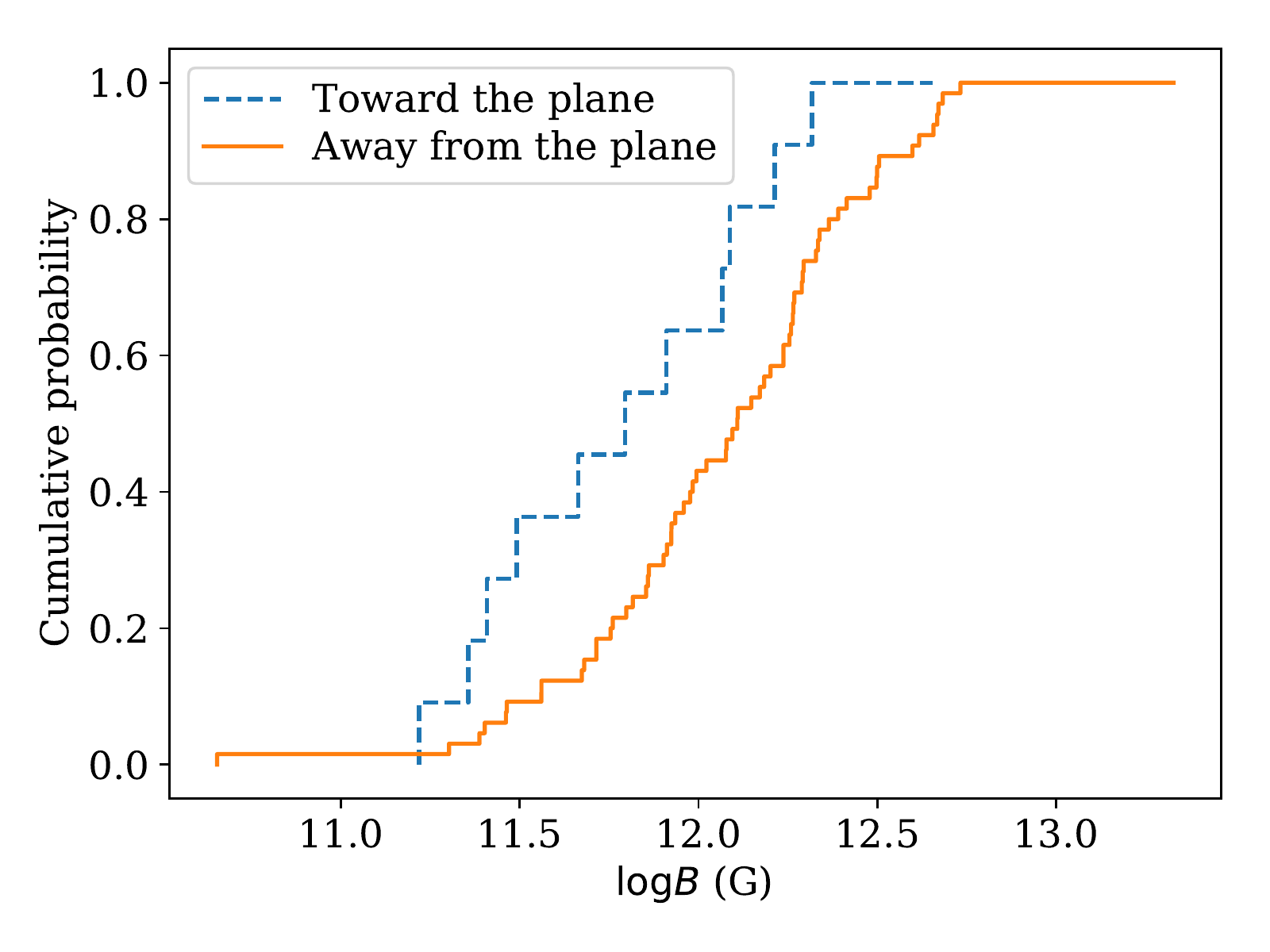}}
\end{minipage}
\caption{Left panel: log-likelihood profile for different values of $t_\mathrm{dec}$ parameters for the complete sample (solid line) and for the sample excluding J0953+0755 (dashed line). The horizontal lines shows the confidence limits (68 per cent and 95 per cent) for $t_\mathrm{dec}$. Right panel: cumulative distribution of magnetic fields for isolated radio pulsars moving toward and away from the Galactic plane. 
\label{f:likel}}
\end{figure*}

When the maximum likelihood approach is used in application to the real sample (Table~\ref{t:res}), we find $t_\mathrm{dec} \sim 12\pm 3$~Myr, see Figure~\ref{f:likel}. The confidence limit is estimated using assumption that $2\log L$ approximately follows $\chi^2$ distribution.
If we remove J0953+0755 from the sample and perform analysis once again, we get very similar result; the log-likelihood function is shifted less than 0.5~Myr.  
The likelihood function grows very fast toward small values of $t_\mathrm{dec}$ rejecting all $t_\mathrm{dec}<8$~Myr with 95 per cent probability. On the other hand, the likelihood function is not as restrictive toward larger values of $t_\mathrm{dec}$. Values of $t_\mathrm{dec} > 20$~Myr (theoretical application limit of the method) are still acceptable with more than 5 per cent probability.
Therefore, our result is the lower limit on the magnetic field decay timescale. The lower limit for the magnetic field decay of 8~Myr translates to upper limit of the crust impurity parameter $Q < 0.25$ following prescription by \cite{2004ApJ...609..999C}.

There is a weak indication that the magnetic field does evolve on timescales comparable to the quarter of the Galactic vertical oscillation period ($\sim 30$~Myr). When cumulative histograms for the magnetic fields $B\propto \sqrt{P\dot P}$ are plotted for pulsars moving away from the plane and toward the plane in Figure~\ref{f:likel}, there is a noticeable shift. Pulsars moving toward the Galactic plane (older) have mean $\log B = 11.8$ while pulsars moving away from the plane (younger in general) have mean $\log B = 12.1$.  

\section{Illustrative cases}
Here we discuss three pulsars which lay too far from the diagonal line of $\tau=t$ in Figure~\ref{f:t_tau}.

\subsection{Curious case of B0950+08}
In a recent study by \cite{pavlov2017} it has been shown that the PSR B0950+08 is too warm for its spin-down age. In the previous analysis by \cite{Noutsos2013} its kinematic age was estimated to be much smaller than the spin-down age. Our analysis reveals the most probable kinematic age to be around 2 Myr, see Figures \ref{f:posterior_tkin_final},\ref{f:t_tau} and \ref{f:gp}. The most important assumption which we made during this analysis is that the pulsar is younger than $\approx 20$ Myr. 

\begin{figure}
\centerline{\includegraphics[width=1\columnwidth]{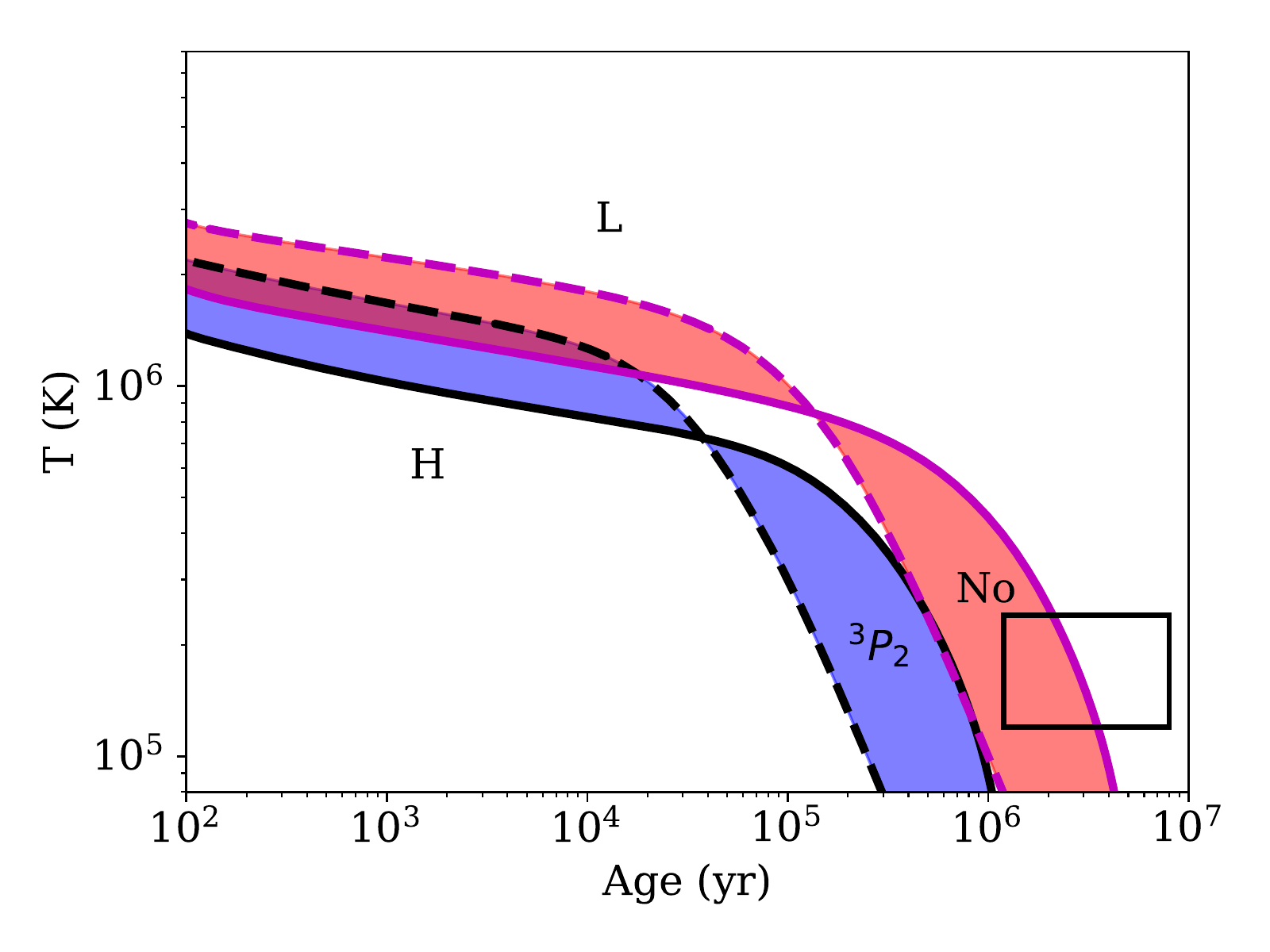}}
\caption{NS cooling curves prepared in framework of the minimal cooling scenario for different compositions of the envelope (H for composition with Fe, L for presence of light elements) as well as with and without $^1P_3$ pairing. The box shows the 68\% credible interval for age and temperature of PSR B0950+08.  
\label{f:temp_age}}
\end{figure}

Our kinematic age estimate shows that the temperature of the pulsar can be explained in framework of the minimal cooling scenario (\citealt{minimalcooling}\footnote{The cooling curves are produced by means of the code \texttt{NSCool} http://www.astroscu.unam.mx/neutrones/NSCool/} maybe with exclusion of special value of $^3P_1$ gap \citealt{baldo1998} 
), see Figure~\ref{f:temp_age}. No additional heating sources are required. 

The age of $t\approx 2$~Myr suggests quite unusual magnetic field evolution such as a fast decay with timescale at once $\tau_\mathrm{dec} \sim 5$ Myr or shorter which is clearly incompatible with other pulsars in our sample.
Another indication of fast magnetic field evolution could be strange values for $\ddot P$ and braking index $n = \nu \ddot \nu / \dot \nu^2$. The braking index of this pulsar computed through second derivative of the frequency is $n\approx -2.3\times 10^3$ \citep{2004MNRAS.353.1311H} which  also might be explained by missed glitches.

The fast magnetic field evolution $t_\mathrm{dec}\sim 5$~Myr can be excluded if the pulsar was born with rotational period which is close to its modern value. This is quite unlikely scenario. To prove it we draw the initial spin-down age distribution based on initial periods and magnetic field distributions from  \cite{faucher2006} and \cite{popov2012}. In both cases only two percent of all outcomes have initial spin down ages which exceed 17~Myr. Thus, the most probable scenario is a combination of some magnetic field decay and longer initial period.  

The origin and evolution of PSR B0950+08 can be better constrained through long timing observations which would allow us to get rid of possible glitches and constrain braking index. In general we expect braking index to be $n>3$ if magnetic field decays or $n=5$ if the magnetic configuration is quadrupole, and $n\ll 0$ if magnetic field grows rapidly as a natural outcome of for e.g. magnetic field re-emergence scenario, see \cite{igoshevpopov2016}. Braking index $n\approx 3$ would mean that the pulsar was most probably born with the values of period and magnetic field  which are close to its modern values.

\subsection{Older than it seems: J0922+0638, J0629+2415}

Another interesting pulsars in our sample are J0922+0638 and J0629+2415 which have the kinematic age three times large than the spin-down age. This could happen if the NS was born from a run-away progenitor or magnetic field increased due to e.g. re-emergence of the magnetic field after fall-back, see \cite*{ho2011, viganopons2012, 2013ApJ...770..106B,igoshevpopov2016}. If the NS progenitor was a part of binary which was disrupted, it could give NS progenitor some speed, so it could on average travel up 130~pc and in rare cases reach 1~kpc distance \citep{2018arXiv180409164R}. The vertical distance of J0922+0638 and J0629+2415 from the plane is 0.72~kpc and 0.32~kpc, so it could be a rare case of binary disruption before radio pulsar was formed. 

The braking index for the first pulsar is large and positive: $n\approx 80$ \citep{2013ApJ...775....2S} which disfavors the magnetic field re-emergence scenario. The reason for large positive braking index is considered to be a sequence of slow glitches. 
The X-ray spectrum of this pulsar shows predominantly non-thermal emission \citep{2015arXiv151107713P} with a possible thermal contribution from a hot polar cap \citep{2018A&A...615A..73R}. This situation is typical for older neutron stars (age more than 1~Myr) which agrees with the kinematic age estimate for this pulsar.

The braking index for the second pulsar is large and negative: $n\approx -210$, so it could be an object with re-emerging magnetic field. On the other hand, the $95$~per cent confidence interval for the kinematic age of J0629+2415 is quite wide and includes the value of its spin-down age.




\section{Discussion: evolution of the pulsar obliquity angle}
The recent MHD simulations by \cite{2014MNRAS.441.1879P} showed that the obliquity angle $\chi$ between the magnetic axis and the rotation axis evolves with time. The obliquity angle determines partly the braking of a pulsar in the case of  plasma filled magnetosphere. The equations for the pulsar braking from \cite{2014MNRAS.441.1879P}
are:
\begin{equation}
\left. \begin{array}{ccc}
P\dot P & = &  (\kappa_0 + \kappa_1 \sin^2\chi) B_p^2 \beta \\
\hspace{0.1cm}\\
\dot \alpha & = & -\kappa_2  \beta \sin \chi \cos \chi  B_p^2/P^2 \\
\end{array} \right\}
\label{e:angle_p}
\end{equation}
with numerical values summarized in Table~\ref{t:num_val}. The value $\beta = \pi^2 R^6 / (c^3 I)$ 
where $I$ is the moment of inertia for the neutron star, $c$ is the light speed and $R$ is the NS radius. 
To fit the example from \cite{2014MNRAS.441.1879P} the value of $\beta$ is chosen to be $\beta = 3\times 10^{-40}$ G s$^{-2}$.
After the system of equations (\ref{e:angle_p}) is solved numerically for $B_p = 10^{12}$~G, initial period of $P_0 = 10$~ms and initial obliquity angle $\chi_0 = 60^\circ$, we compute the spin down age $\tau_\mathrm{mod} = P/(2\dot P)$ for each real age $t$ based on period and period derivative. The dependence $\tau_\mathrm{mod}$ vs. $t$ is shown in Figure~\ref{f:t_tau}. This line does not differ much from $\tau = t$. 
Therefore this model agrees with the observed sample of the radio pulsars.


\section{Conclusion}

We derive the posterior probability densities for three dimensional velocities of radio pulsars. These values can be especially useful in analysis of the millisecond radio pulsar ensemble to correct for the Shklovskii effect.
We suggest a new Bayesian estimate for kinematic ages of radio pulsars with the spin-down age $\tau < 20$~Myr. This estimates takes into account the bimodality of the velocity distribution shown in \cite{verbunt2017} and uncertainty in distance and proper motion measurements.

According to the new estimate, the lower limit on the exponential magnetic field decay timescale is $8$~Myr. The maximum likelihood estimate gives slight preference for decay timescale $t_\mathrm{dec}\approx 12$~Myr, and larger magnetic field decay timescales (even > 20~Myr) are compatible with observations.
Absence of magnetic field decay on $1-20$~Myr timescale does not contradict results about moderate magnetic field decay identified in \cite{2014MNRAS.444.1066I} since that decay occurs much earlier ($\tau < 1$~Myr) and it stops afterwards. 

In the case of J0953+0755 two factors seem to play a role: magnetic field decay and longer initial period. The kinematic and cooling ages of J0953+0755 are $\approx 2$~Myr while its spin-down age is $17$~Myr which is order of magnitude larger. 
There are multiple explanations for this strange behavior: (1) hidden heating sources, (2) large initial rotational period and (3) complicated  magnetic field evolution. The first hypothesis is adhoc and does not explain the small kinematic age of the pulsar. The second hypothesis is unlikely to be solely responsible for this discrepancy, we showed that such a combination of large initial rotational period and small magnetic field occurs in $\approx2$ percent of cases. The third hypothesis is the most probable one, since it naturally explains both strange braking index and coincidence of kinematic and cooling ages (these ages do not depend on magnetic field evolution). Complicated magnetic field evolution can be a consequence of high impurity of the inner crust. Further studies of the PSR B0950+08 (timing and X-ray) are highly desirable to better understand its unusual properties.


\section*{Acknowledgements}
      A.I. thanks Frank Verbunt, Sergei Popov and Buscicchio Riccardo for many fruitful discussions. A.I. is grateful for a chance to participate in the Astro Hack Week 2018 hosted by the Lorentz center. A.I. acknowledges support from the Israel science foundation I-CORE program 1829/12.

\bibliographystyle{mnras} 
\bibliography{msp}

\appendix
\section{Numerical calculations of integral eq. (3)}
\label{a:first}
Both integrals in the nominator of eq.(\ref{e:poster_vD}) are impossible to compute analytically because of the exponent depending on sine and cosine of the velocity orientation angles. The numerical integration is quite challenging to perform because the joint probability peaks sharply when angles $\xi_1$ and $\xi_2$  are similar to the orientation of the measured proper motion and ratio of velocity to distance is similar to  the measured length of the proper motion vector.

To deal with this difficulty we introduce two auxiliary angles:
\begin{equation}
\tan \xi_{2m} = \frac{\mu_{\alpha *}' - \mu_{\alpha *, G}(D)}{\mu_\delta ' - \mu_{\delta ,G}(D)}
\label{e:xi_2m}
\end{equation}
which determines the preferable orientation on the plane of sky. The values $\mu_{\alpha *}(D)$ and $\mu_{\delta ,G}(D)$ are the distance dependent correction for the motion of the local standard of rest. 
The angle in eq. (\ref{e:xi_2m}) was used in \cite{verbunt2017}. The second angle is:
\begin{equation}
\sin \xi_{1m} = \frac{(\mu_\delta ' - \mu_{\delta ,G}(D))D }{v \sin \xi_{2m}}
\end{equation}
This angle is not determined when $|(\mu_\delta ' - \mu_{\delta ,G}(D))D / (v \sin \xi_{2m})| > 1$ which indicates that the magnitude of the velocity is not enough to reproduce the measured proper motion.
Both integrals in eq. (\ref{e:poster_vD}) are computed in three sub-intervals with different numerical step:
\begin{equation}
\int_0^a f(x) dx = \int_0^{\xi_m - h} f(x) dx + \int_{\xi_m -h }^{\xi_m+h} f(x) dx + \int_0^{a} f(x) dx
\end{equation}
where $f(x) dx$ stands for function $P_\mathrm{sim}$. The limits $a$ is $\pi$ in the case of integration over $\xi_1$ and $2\pi$ in the case of the integration over $\xi_2$. The value of $h$ is fixed at value $\pi / 70.0$ which resolves the sharp peak efficiently.

\section{Integration over initial Galactic height}
\label{a:thin}
We rename the auxiliary variables eqs.~(\ref{e:a1}, \ref{e:b1} and \ref{e:c1}):
\begin{equation}
A_1 = \frac{1}{2\sigma^2} + \frac{1}{2\sigma_b^2 D^2} + \frac{\cot^2 b}{2\sigma^2}
\end{equation}
\begin{equation}
A_2 = \frac{\mu_{b, G} - \mu_b'}{D\sigma_b^2} - \frac{\cot b (D - z_0/\sin b)}{t\sigma^2}
\end{equation}
\begin{equation}
A_3 = \frac{(\mu_{b, G} - \mu_b')^2}{2\sigma_b^2} + \frac{(D - z_0 / \sin b)^2}{2t^2\sigma^2}
\end{equation}
The variables $A_2$ and $A_3$ include the $z_0$, while the variable $A_1$ does not. It allows us to rewrite the expression in form of the second degree polynomial:
\begin{equation}
\exp\left(\frac{A_2^2}{4A_1} - A_3 - \frac{|z_0|}{h}\right) = \exp\left( -B_1 z_0^2 - B_2^{(+,-)}z_0 - B_3\right)
\end{equation}
with new auxiliary variables:
\begin{equation}
B_1 = \frac{1}{2t^2\sin^2 b \sigma^2} - \frac{\cot^2 b}{4A_1t^2 \sin ^2 b \sigma^4}
\end{equation}
\begin{equation}
B_2^{(-)} = - \frac{(\mu_{b, G} - \mu_b')\cot b}{2A_1 t\sin b \sigma^2 \sigma_b^2 D} + \frac{D\cot^2 b}{2A_1 t^2 \sigma^4 \sin b} - \frac{D}{\sigma^2 t^2 \sin b} - \frac{1}{h_0}
\end{equation}
\begin{equation}
B_2^{(+)} = - \frac{(\mu_{b, G} - \mu_b')\cot b}{2A_1 t\sin b \sigma^2 \sigma_b^2 D} + \frac{D\cot^2 b}{2A_1 t^2 \sigma^4 \sin b} - \frac{D}{\sigma^2 t^2 \sin b} + \frac{1}{h_0}
\end{equation}
\begin{equation}
B_3 = \frac{D^2}{2\sigma^2 t^2} + \frac{(\mu_{b,G}-\mu_b')^2}{2\sigma_b^2} - \left[\frac{(\mu_{b,G} - \mu_b')}{D\sigma_b^2} -\frac{D\cot b}{t\sigma^2}\right]^2\frac{1}{4A_1}
\end{equation}
The reason to introduce two separate $B_2^{(+,-)}$ variables is that the scale height distribution depends on absolute value of $z_0$ and not on $z_0$ itself. It makes us to split integral into a sum of integrals, one from $-\infty$ to 0 and another one from $0$ to $\infty$.
After the integration over $z_0$ the equation contains terms:
\begin{equation}
\chi^- =  \left\{1 + \mathrm{erf} \left( \frac{B_2^{(-)}}{2\sqrt{B_1}} \right) \right\}\exp \left( \frac{B_2^{(-)2}}{4B_1} - B_3\right)
\end{equation}
\begin{equation}
\chi^+ =  \left\{1 - \mathrm{erf} \left( \frac{B_2^{(+)}}{2\sqrt{B_1}} \right) \right\}\exp \left( \frac{B_2^{(+)2}}{4B_1} - B_3\right)
\end{equation}
This term is especially difficult to compute when $x=B_2^{(-)} / (2\sqrt{B_1}) < 0$ and large, in this case $1+\mathrm{erf} (x) \to 0$ and $\exp(x^2 - B_3) \to \infty$. 
To deal with this difficulty we use the asymptotic expansion for the error function for negative values of $x$:
\begin{equation}
\chi^-  = -\exp(-B_3)\frac{1}{\sqrt{\pi}x}\left[1 - \frac{1}{2x^2} + \frac{3}{4x^4} -\frac{15}{8x^6} +
... \right]
\end{equation}
\begin{equation}
\chi^+  = \exp(-B_3)\frac{1}{\sqrt{\pi}x}\left[1 - \frac{1}{2x^2} + \frac{3}{4x^4} -\frac{15}{8x^6} +
... \right]
\end{equation}

The result of integration is written as:
\begin{equation}
p(t | D) = \frac{h_0}{4\sigma^2 t^2 \sqrt{A_1 B_1}} \left[D (\chi^- + \chi^+) + \frac{1}{2B_1 \sin b}\left( B_2^{(-)} \chi^-  + B_2^{(+)} \chi^+ \right) \right] 
\end{equation}

\bsp	
\label{lastpage}
\end{document}